\documentclass[reprint,
superscriptaddress,
 amsmath, amssymb, aps, pra,
floatfix,
]{revtex4-2}
\usepackage{blindtext}
\usepackage{graphicx}
\usepackage{dcolumn}
\usepackage{bm}
\usepackage{mathtools}
\usepackage{gensymb}
\usepackage{amsmath,physics}
\usepackage{calrsfs}
\DeclareMathAlphabet{\pazocal}{OMS}{zplm}{m}{n}
\usepackage{qcircuit}
\usepackage{dsfont,soul,xcolor}
\usepackage[caption=false]{subfig}
\usepackage[export]{adjustbox}
\makeatletter
\newcommand*{\addFileDependency}[1]{
  \typeout{(#1)}
  \@addtofilelist{#1}
  \IfFileExists{#1}{}{\typeout{No file #1.}}
}
\makeatother

\DeclareMathSymbol{\shortminus}{\mathbin}{AMSa}{"39}

\usepackage{amsthm}
\theoremstyle{plain}

\begin{document}


\title{Bose-Marletto-Vedral experiment without observable spacetime superpositions}
\author{Nicetu Tibau Vidal}
\affiliation{QICI Quantum Information and Computation Initiative, Department of Computer Science, The University of Hong Kong, Pok Fu Lam Road, Hong Kong}%

\affiliation{HKU-Oxford Joint Laboratory for Quantum Information and Computation}

\author{Chiara Marletto}
\affiliation{%
Clarendon Laboratory, Department of Physics, University of Oxford, Oxford OX1 3PU, United Kingdom
}%

\author{Vlatko Vedral}
\affiliation{%
Clarendon Laboratory, Department of Physics, University of Oxford, Oxford OX1 3PU, United Kingdom
}%

\author{Giulio Chiribella}
\affiliation{QICI Quantum Information and Computation Initiative, Department of Computer Science, The University of Hong Kong, Pok Fu Lam Road, Hong Kong}%

\affiliation{HKU-Oxford Joint Laboratory for Quantum Information and Computation}

\begin{abstract}
Reconciling quantum mechanics and general relativity remains one of the most profound challenges in modern physics. The BMV (Bose-Marletto-Vedral) experiment can assess the quantum nature of gravity by testing whether gravitational interactions can generate entanglement between quantum systems. 
In this work, we show that entanglement can be generated by gravity without requiring spacetime superpositions or quantum spacetime degrees of freedom by using mediators that do not satisfy the usual property of local tomography when coupling to quantum matter. Specifically, we showcase how entanglement can be generated using three distinct toy models that display non-locally tomographic couplings between quantum matter and a locally classical gravitational mediator. These models include (i) fermionic systems with the parity superselection rule, (ii) non-Abelian anyonic systems, and (iii) a novel bit anti-bit model. 
Our results demonstrate a crucial point: a gravitational mediator which does not exhibit superpositions of its classical basis but still qualifies as non-classical via non-locally tomographic coupling mechanisms can generate entanglement through local interactions. This work also underscores the importance of relaxing local tomography in exploring the quantum-gravitational interface. It provides a novel perspective on the role of spacetime degrees of freedom in entanglement generation through local interactions.
\end{abstract}

\maketitle


\section{Introduction} \label{sec:intro}


The Bose-Marletto-Vedral (BMV) or gavitationally induced entanglement (GIE) experiment \cite{marletto_quantum-information_2025,marletto_gravitationally_2017,marletto_why_2017, marletto_witnessing_2020, bose_spin_2017,martin-martinez_what_2023,christodoulou_locally_2023,bengyat_gravity-mediated_2024, galley_any_2023,krisnanda_observable_2020, di_pietra_bose-marletto-vedral_2024,di_pietra_role_2024,weber_bose-marletto-vedral_2024,kent_should_2024,kent_small_2024,kent_testing_2021,hidaka_entanglement_2023,zhang_entanglement_2025,li_generation_2023,mari_can_2025,yant_operational_2025,van_manen_causal_2025,feng_collapse-based_2025,schut_relaxation_2023,telali_causality_2025,streltsov_quantum_2024} serves as a pivotal platform for exploring the quantum features of gravity, aiming to test whether gravitational interactions can create entanglement between quantum systems. Traditional approaches to understanding locality in quantum mechanics often rely on the assumption of local tomography, which states that the global state of a system can be reconstructed from local measurements. However, this assumption may not hold in strongly gauged systems, such as gravitational interaction.

In this paper, we investigate the consequences of relaxing the local tomography requirement for the gravitational mediator in the BMV experiment. We show that gravitational mediators can mediate entanglement generation between quantum systems while also violating the assumption of local tomography. In particular, the non-classical degrees of freedom of the gravitational mediator may not be locally single-shot measurable, nor may they be preparable.  By analysing three distinct toy models embodying different physical principles, we demonstrate that the entanglement mediator may be non-classical while also not being preparable in a superposition of its classical basis states.  This work showcases how, in spacetime with no preparable superposition, entanglement generation through local interaction can still occur. We pinpoint the role of non-locally tomographic couplings as the culprit for this phenomenon. We expand on how these couplings can appear in models where general relativity is coupled to quantum matter, given that such interaction may be subject to superselection rules. 

The structure of the paper is as follows. We introduce local tomography in Section \ref{sec:loctom}. We explain the role of this property in existing models of the BMV effect, stressing how it is not a necessary assumption. We further present and explain how non-locally tomographic couplings are expected in constrained quantum systems. Section \ref{sec:localclass} discusses three specific examples where we use known non-locally tomographic theories to explicitly showcase how entanglement can be generated through local interaction with a locally classical mediator. The models we use are fermions in Subsection \ref{subsec:fermions}, non-Abelian Ising anyons in Subsection \ref{subsec:nonabelian} and the bit and anti-bit framework in Subsection \ref{subsec:bitanti}. We conclude with a discussion in Section \ref{sec:discuss}, where we comment on and explore the implications of our results. 




\begin{figure}[bh!]
    \centering
    \includegraphics[width=\linewidth]{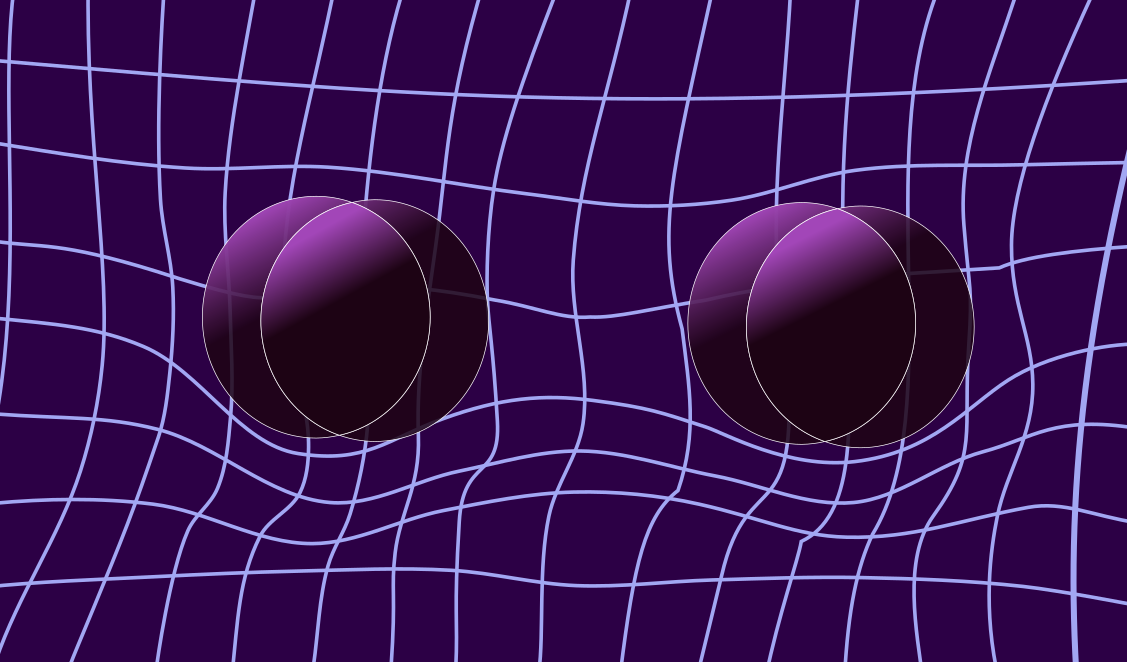}
    \caption{Two masses each in a position superposition can be expected to get entangled by system local gravitational interaction mediated by spacetime itself. The BMV arguments are used to infer that spacetime itself cannot be a classical system.}
    \label{fig:art}
\end{figure}

\section{Local tomography in BMV}\label{sec:loctom}

The BMV experiment \cite{marletto_gravitationally_2017, marletto_why_2017, bose_spin_2017} proposes that by measuring the capability of generating entanglement between two quantum systems by gravitational interaction, one can probe the presence of quantum properties in gravity. Different arguments have been proposed within different frameworks and assumptions, with slightly different consequences. Nevertheless, the key ideas of these arguments can be summarised as follows. 

Consider two quantum systems $Q_1, Q_2$ (assume, for simplicity, that they are qubits), and a gravitational/spacetime system $M$ that mediates the interaction between $Q_1, Q_2$. Let us assume that the sector $M$ is classical - by this we shall mean, following the general approaches in \cite{marletto_witnessing_2020,galley_any_2023}, that $M$'s pure states are all single-shot distinguishable and they compose following the usual tensor product composition. 

Under the global composite system $Q_1 M Q_2$, consider only {\sl local} interactions in $Q_1 M$ and in $M Q_2$, which we define as interactions such that $M$ can directly interact with $Q_1$ and with $Q_2$ separately, but $Q_1$ and $Q_2$ cannot interact with one another. Then upon preparing a separable quantum state on $Q_1, Q_2$, by applying local interactions in the global system $Q_1 M Q_2$ and discarding $M$, a separable state is obtained on $Q_1 Q_2$. Hence, no entanglement can be generated between $Q_1$ and $Q_2$ by this local mediation.  This result (called the General Witness Theorem (GWT), \cite{marletto_quantum-information_2025}) can also be proven in a general information-theoretic context, without using the formalism of quantum theory, \cite{marletto_witnessing_2020}, which makes the BMV experiment rest on a robust theoretical basis. Here we shall demonstrate how robust it is, by modelling the BMV experiment with non-standard quantum systems, which violate the assumption of local tomography. 

Interestingly we showcase how the violation of local tomography couplings is enough to mediate entanglement and thus violate classicality. We showcase the non-classicality of systems that appear to be locally classical but do not couple via the usual tensor product mechanism. The mediators in our models do not contain any non-commuting observables nor superpositions; but they are still able to mediate entanglement.    

First, let us briefly summarise what local tomography is.   
In a theory that contains different subsystems $A, B, C,\dots$, the property of local tomography consists in requiring that by coordinating and using localised measurements and actions in the subsystems, one can reconstruct the global state of the system \cite{hardy_theory_2013, hardy_reconstructing_2016, hardy_quantum_2001, dariano_fermionic_2014}. Within the context of bipartite regular quantum theory, local tomography is codified by seeing that two states $\rho_{AB}=\sigma_{AB}$ are the same if and only if $\Tr\left(\hat{M}_A\otimes \hat{N}_B \rho_{A B}\right)=   \Tr\left(\hat{M}_A\otimes \hat{N}_B \sigma_{A B}\right)$ for all local observables $\hat{M}_A, \hat{N}_B$. Digging a bit deeper, we see that the property that ensures this equivalence in the case of bipartite regular quantum theory is its linearity and the fact that any global observable can always be decomposed as a linear combination of tensor products of local observables as
\begin{eqnarray}
    \hat{O}_{AB}=\sum_{ij} o_{ij} \hat{M}^{(i)}_A \otimes \hat{N}^{(j)}_B . 
\end{eqnarray}

Local tomography is a property that has been extended to general physical theories \cite{barnum_generalized_2007, barrett_information_2007}, and it is regularly used as an assumption when discussing physical systems and quantum foundations \cite{araki_characterization_1980, barrett_information_2007, barnum_generalized_2007, hardy_quantum_2001, chiribella_probabilistic_2010, chiribella_informational_2011, chiribella_process_2021}. It is an established relevant property of physical theories that we should consider whether it applies to our modelling scenarios and the consequences it carries. 

We shall now highlight how local tomography is not a key assumption of the GWT and it is not necessary to model the BMV experiment: we do not need to assume it in order to conclude that a local mediator of entanglement is non-classical. However, local tomography has so far been used in various models of the BMV effect. For instance, in  \cite{bose_spin_2017} it is assumed when modelling the gravitational interaction using linearised quantum gravity, where the algebra of observables of the matter-gravity system is given by the free algebra spanned by the products of the matter and gravitational fields and conjugate field momenta. In \cite{marletto_gravitationally_2017}, it is assumed implicitly when stating the general form of a state in the composite system $Q_1 M Q_2$, where we can see how the state is considered as a normalised global observable, which it is assumed to be linearly decomposable in terms of the products of local observables of $Q_1 Q_2$ and $T$. Moreover, in GPT analysis of the BMV experiment, such as \cite{galley_any_2023}, the allowed quantum-classical compositions satisfy local tomography by definition. 


One could argue that local tomography is a natural property required by any physical theory. In some approaches, it even appears as one of the axioms of quantum theory \cite{hardy_quantum_2001, hardy_reconstructing_2016, chiribella_informational_2011, masanes_derivation_2011, masanes_existence_2013, barnum_higher-order_2014, wilce_conjugates_2019,selby_reconstructing_2021,dakic_quantum_2011}. Still, two important questions arise. First, do we expect local tomography to be satisfied when coupling gravity and matter? Second, if we consider a theory of gravity-matter interaction that does not satisfy local tomography, what physical consequences appear in the context of the BMV experiment? To answer these two key questions, we now explore known non-locally tomographic composite theories. 
 
\subsection{Non-locally tomographic couplings}\label{subsec:loctomcoup}

Let us take the point of view that, given two physical theories that govern two different sectors, $Q$ and $M$, we want to construct a global composite non-locally tomographic theory, $ QM$. Within this context, we can discuss how we couple the two sectors. Therefore, we refer to the fact that local tomography is a property of the global overarching theory, not of the local theories that rule independently in the sectors $Q$ and $M$. To remark on this position, instead of talking about non-local tomographic theories, we explicitly talk about non-locally tomographic couplings between two sectors.

A result we use throughout this work is that couplings of constrained quantum systems are expected to be non-locally tomographic \cite{dariano_fermionic_2014}. Consider a finite-dimensional non-constrained quantum sector $A$ and a constrained quantum sector $B$. $N_A=\dim{A}^2$ and $N_B<\dim{B}^2$ are the number of linearly independent local observables in $A$ and $B$ respectively. If we can couple $A$ and $B$ in such a way that $N_{AB}> N_A N_B$, we expect to have non-local tomography since at least one of the global quantum observables will not be linearly decomposable as products of local observables. 

Let us consider a simple constrained quantum system where the number of linearly independent observables in a system scales as $N_A =\frac{\dim{A}^2}{2}$. Then, we obtain that if we consider a coupling of two such subsystems where $\dim{A A'}=\dim{A} \dim{A'}$, then $N_{A A'}=\frac{\dim{A A'}^2}{2}= 2 N_A N_{A'}> N_A N_{A'}$, obtaining the non-local tomography property. Such a simple example is satisfied in the case of fermions under the parity superselection rule (SSR) \cite{dariano_fermionic_2014, dariano_feynman_2014, vidal_local-realistic_2022, vidal_quantum_2021} that we later use as an explicit toy example. 

We must focus on such constrained systems because the puzzle of coupling gravity and matter leads us to it. Essentially, we aim to investigate the effects of a classical theory of gravity combined with a quantum theory of matter. Since classical theory can be regarded as strongly constrained quantum theory \cite{sherry_interaction_1978}, we are drawn directly to the possibility of non-locally tomographic couplings of an unconstrained quantum system (matter) and a constrained quantum system (gravity). 

Another reason it is relevant to focus on constrained quantum systems is that, generally, gauge field theories in quantum theory are constrained. In such theories, not all Hermitian operators are physical observables. Instead, only the gauge invariant Hermitian operators can be considered physical observables \cite{giesel_introduction_2008, chataignier_construction_2020}. We believe it is reasonable to expect that the coupling of quantum matter with gravity can be modelled as a gauge field theory with a quantum sector. Thus, we believe there are reasonable arguments that prevent us from considering the coupling between the gravitational and matter sectors to be locally tomographic.

\section{Mediating entanglement using locally-classical mediators}\label{sec:localclass}

We now turn our attention to the second key question we aim to address. If we have a non-locally tomographic coupling, what physical implications does it have in the BMV experiment scenario? 

We show how entanglement between the quantum two-level systems $Q_1, Q_2$ can be mediated by a system $M$ with a single non-trivial local observable $T$. In the following section, we explicitly show such mediation using three different non-locally tomographic toy theories that model the quantum and mediating systems. 

\subsection{Fermions} \label{subsec:fermions}

For our first example, we use fermionic theory, which incorporates a superselection rule (SSR) \cite{wick_intrinsic_1952, friis_reasonable_2016, friis_fermionic-mode_2013, vidal_quantum_2021, vidal_local-realistic_2022}. Let us consider a system of $5$ spinless fermionic modes for the BMV scenario. We will use modes $1,2$ to encode the qubit $Q_1$, modes $4,5$ to encode the qubit $Q_2$, and mode $3$ to represent the mediator $M$. 

Fermions are fundamental particles that obey a specific anticommutation algebra of creation and annihilation operators. To not violate the no-signalling principle \cite{barnum_generalized_2007, ghirardi_general_1980, ghirardi_experiments_1988}, fermions must be constrained under the parity superselection rule (SSR) \cite{wick_intrinsic_1952, vidal_quantum_2021}. The fermionic SSR algebra is constructed using fermionic creation and annihilation operators, 
\begin{eqnarray}
    \hat{f}_j \ket{0}=0, \quad \{\hat{f}_j,\hat{f}_k\}=0, \quad \{\hat{f}_j,\hat{f}_k^\dagger\}=\delta_{j k}  \mathbb{I}
\end{eqnarray}
where $\ket{0}$ is the vacuum state and each $\hat{f}_j$ is the fermionic annihilation operator on mode $j$. Significantly, fermionic annihilation operators anticommute, thus $\hat{f}_j ^2 =0$.  The parity SSR imposes that the only physically allowed observables are the Hermitian operators given by polynomials of the annihilation and creation operators such that each monomial has an even degree. In other words, $\hat{f}_j \hat{f}_j^\dagger, \hat{f}_j\hat{f}_k+\hat{f}_k^\dagger \hat{f}_j^\dagger$ are physical observables, but $\hat{f}_j + \hat{f}_j^\dagger$ or $\hat{f}_j^\dagger \hat{f}_j \hat{f}_k+\hat{f}_k^\dagger \hat{f}_j^\dagger \hat{f}_j$ are not despite being Hermitian fermionic operators.  

Under this restriction, the number of linearly independent fermionic SSR physical observables scales as $N_n=2^{2 n-1}$ for $n$ fermionic modes. This contrasts with the unrestricted case where the number of linearly independent Hermitian operators is $2^{2n}$. Based on the argument we have presented before, we know the SSR fermion's coupling is non-locally tomographic. The SSR restriction has been shown to be necessary to have the field locality condition of microcausality that ensures that the no-signalling principle is satisfied \cite{vidal_quantum_2021}. This means that any physical local observable in $Q_1$ commutes with any physical local observable in $Q_2$ or $M$ and vice versa. 

The only linearly independent physical local observables for mode $j$ are $\hat{f}_j \hat{f}_j^\dagger$ and $\hat{f}_j^\dagger \hat{f}_j$. We have $\mathbb{I}=\hat{f}_j^\dagger \hat{f}_j+\hat{f}_j \hat{f}_j^\dagger$ and define $T_j=\hat{f}_j \hat{f}_j^\dagger-\hat{f}_j^\dagger \hat{f}_j$. From this, we can see how the SSR global observable $\hat{f}_j \hat{f}_k +\hat{f}_k^\dagger \hat{f}_j^\dagger$ cannot be decomposed as a linear combination of products of local observables in $j$ and $k$. Moreover, all local observables in each mode $j$ commute, and the local algebra in mode $j$ can be generated from the single observable $T_j$. 

Therefore, in our BMV scenario, the mediator $M$ identified with the fermionic mode $3$ locally behaves like a classical bit. It has a single local generator observable $T_3$. Therefore, an observer in $M$ would only be able to locally measure and prepare as if their local system were a classical bit. Despite these local classical features in $M$, we show how entanglement between $Q_1$ and $Q_2$ can be locally mediated through $M$.    

\begin{figure}[b]
    \centering
    \includegraphics[width=\linewidth]{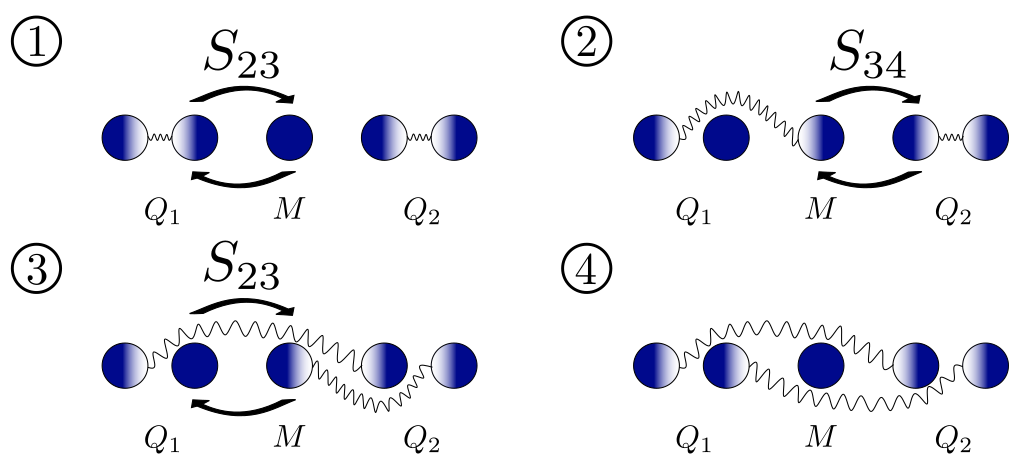}
    \caption{Figure displaying the entanglement mediation using the fermionic SSR algebra as the toy model. The connected half-filled circles represent the $\ket{+}$ superposition states between those modes.}
    \label{fig:fermion}
\end{figure}

$Q_1$ and $Q_2$ can be considered quantum systems since they hold their respective non-commuting local observables $\hat{Z}_1 \equiv \hat{f}_1^\dagger \hat{f}_1 \hat{f}_2 \hat{f}_2^\dagger $, $\hat{X}_1 \equiv \hat{f}_1^\dagger \hat{f}_2 +\hat{f}_2^\dagger \hat{f}_1 $ for $Q_1$ and $\hat{Z}_2 \equiv \hat{f}_4^\dagger \hat{f}_4 \hat{f}_5 \hat{f}_5^\dagger $, $\hat{X}_2 \equiv \hat{f}_4^\dagger \hat{f}_5 +\hat{f}_5^\dagger \hat{f}_4 $ for $Q_2$. Similarly to the BMV experiment, let us prepare $Q_1$ and $Q_2$ in equivalent $\ket{+}_i$ states. Furthermore, let us prepare $M$ in the state equivalent to $\ket{1}$. See Figure \ref{fig:fermion} for a diagrmaatic representation of the entanglement mediation protocol. 

The fermionic initial state we consider is  
\begin{eqnarray}
    \ket{\psi_0}=\frac{1}{2}\left(\hat{f}_1^\dagger+\hat{f}_2^\dagger\right) \hat{f}_3^\dagger\left(\hat{f}_4^\dagger+\hat{f}_5^\dagger\right) \ket{0}.
\end{eqnarray}

Notice it is a separable pure product state in any bipartition of $Q_1 M Q_2$. Importantly, though, we need to use the fermionic partial trace \cite{friis_reasonable_2016, vidal_quantum_2021}, 
 \begin{gather}
    \Tr_j\left( \hat{f}_1^{\dagger ^{s_1}} \cdots  \hat{f}_j^{\dagger^{s_j}} \cdots \hat{f}_n^{\dagger^{s_n}} \ketbra{0}{0}   {\hat{f}_n}^{r_n} \cdots {\hat{f}_j}^{r_j} \cdots {\hat{f}_1}^{r_1} \right) = \nonumber\\ \scalebox{0.87}{$\delta_{s_j r_j} (-1)^{\sum_{k=j+1}^{n} (s_j s_k+ r_j r_k)}   \hat{f}_1^{\dagger ^{s_1}} \cdots  \cdots \hat{f}_n^{\dagger^{s_n}} \ketbra{0}{0}   {\hat{f}_n}^{r_n} \cdots \cdots {\hat{f}_1}^{r_1}$} 
\end{gather}

which correctly discards fermionic modes considering the fermionic algebra. In the expression, $s_i, r_i \in \{0,1\}$, whether the mode is in the vacuum or excited. The fermionic partial trace just anticommutes the creation and annihilation operators of the trace mode towards the vacuum, and then applies the usual partial trace. Using it to discard $M$, it is straightforward to check that the reduced quantum state for $ Q_1Q_2$ is a separable pure product state between $Q_1 | Q_2$ with the expression
\begin{eqnarray}
    \ket{\psi_0^{12}}=\frac{1}{2}\left(\hat{f}_1^\dagger+\hat{f}_2^\dagger\right)\left(\hat{f}_4^\dagger+\hat{f}_5^\dagger\right) \ket{0}
\end{eqnarray}

We know that this pure state above is separable, since it is an uncorrelated state. In other words is a state that satisfies
\begin{gather}
      \Tr(\hat{O}_{Q_1}   \ketbra{\psi_0^{1 2}}  )\cdot \Tr(\hat{O}_{Q_2} \ketbra{\psi_0^{1 2}} ) \nonumber \\ =\Tr(\hat{O}_{Q_1} \cdot \hat{O}_{Q_2} \ketbra{\psi_0^{1 2}}) \label{eq:uncorr}
\end{gather}
for all physically local observables $\hat{O}_{Q_1}$ and $\hat{O}_{Q_2}$. 

Therefore, our initial state of $Q_1 | Q_2$ is a separable state. Let us now apply the entanglement generation protocol. We use unitaries that are local in $Q_1 M$ or $M Q_2$ so the mediation of entanglement can be considered to be system local \cite{biagio_circuit_2025}. Under a locally tomographic coupling, we could not see the systems $Q_1 Q_2$ getting entangled as shown in one of the BMV papers \cite{marletto_gravitationally_2017}. However, in this fermionic setting, we know local tomography is not satisfied. The local mediating unitaries we choose are just fermionic swap gates. Such unitaries can be defined by their effect on the fermionic annihilation operators. $S_{ij}$ swaps the fermionic annihilation operator of the mode $i$ for the one in $j$ as
\begin{eqnarray}
    S_{ij} \hat{f}_k S_{ij}^\dagger= \begin{cases}
        \hat{f}_k & \text{if} \medspace k \neq i \medspace \& \medspace  k \neq j \\
         \hat{f}_j & \text{if } k =i   \\
        \hat{f}_i & \text{if } k=j \\
    \end{cases} 
\end{eqnarray}

Since $S_{2 3}$ does not involve or impact in any way $Q_2$, and $S_{34}$ does not do so in $Q_1$, both qualify as system local unitaries. Now, let us consider the resulting state of alternating such system local unitaries in the following way as in Figure \ref{fig:fermion}
\begin{eqnarray}
     S_{23} S_{34} S_{23} \ket{\psi_0} &=  S_{23} S_{34} \frac{1}{2}\left(\hat{f}_1^\dagger+\hat{f}_3^\dagger\right) \hat{f}_2^\dagger\left(\hat{f}_4^\dagger+\hat{f}_5^\dagger\right) \ket{0} \nonumber\\ &= S_{23}  \frac{1}{2}\left(\hat{f}_1^\dagger+\hat{f}_4^\dagger\right) \hat{f}_2^\dagger\left(\hat{f}_3^\dagger+\hat{f}_5^\dagger\right) \ket{0} \nonumber \\& = \frac{1}{2}\left(\hat{f}_1^\dagger+\hat{f}_4^\dagger\right) \hat{f}_3^\dagger\left(\hat{f}_2^\dagger+\hat{f}_5^\dagger\right) \ket{0}
\end{eqnarray}

In the above manipulation we use $S_{j k}^\dagger = S_{jk}$ and $S_{jk} \ket{0}=\ket{0}$. Taking the fermionic partial trace of mode $3$ to eliminate the mediator sector $M$, we obtain the final reduced state for $Q_1 Q_2$. The state
\begin{eqnarray}
    \ket{\psi_f^{12}}=\frac{1}{2}\left(\hat{f}_1^\dagger+\hat{f}_4^\dagger\right)\left(\hat{f}_2^\dagger+\hat{f}_5^\dagger\right) \ket{0}
\end{eqnarray}
is pure, however, by using fermionic information theory we can see it is in fact maximally entangled in the bipartition $Q_1 | Q_2$. We can see this by noting that it is a pure state and considering the reduced states in $Q_1$ and $Q_2$, observing that they are the maximally mixed states 
\begin{eqnarray}
    \rho_{Q_1}=\frac{1}{4} \mathbb{I}_{Q_1}  \qquad   \rho_{Q_2}=\frac{1}{4} \mathbb{I}_{Q_2}  
\end{eqnarray}

Indeed, then we can find the expectation values of the local SSR observables $\hat{X}_1$ and $\hat{X}_2$ that are the embedded qubit Pauli $x$ operators give us
\begin{eqnarray}
    \Tr(\hat{X}_1 \rho_{Q_1})= \Tr(\hat{X}_2 \rho_{Q_2})= 0 \\ \Tr(\hat{X}_1 \cdot \hat{X}_2 \ketbra{\psi_f^{1 2}}) =-\frac{1}{2}
\end{eqnarray}
therefore showcasing that the state $\ket{\psi_f^{1 2}}$ is pure and not uncorrelated, thus entangled.

Therefore, we have seen how by applying system local unitaries $S_{23}$ and $S_{34}$, $M$ can locally mediate the generation of entanglement between the two quantum sectors $Q_1$ and $Q_2$ even though $M$ has a single observable generator $T_3$. Such behaviour is only possible under non-locally tomographic couplings. 

A way to interpret this is that when $M$ is coupled to $Q_1$ and $Q_2$ in this non-trivial way, the global system can access some hidden degrees of freedom that are forbidden to be accessed when considering $M$ locally. To further convince the reader that $M$ is behaving locally like a classical bit, let us consider the reduced state for $M$ at each stage of the mediation. 

It is straightforward to see that the expressions of $\rho_M(t)$, calculated by applying the fermionic partial trace of modes $1,2,4$ \& $5$, when represented in the $\{\ket{0}, \hat{f}_3^\dagger \ket{0}\}$ basis follow the sequence
\begin{eqnarray}
    \begin{pmatrix}
        0 & 0 \\ 0 & 1
    \end{pmatrix}, \quad \begin{pmatrix}
        \frac{1}{2} & 0 \\ 0 & \frac{1}{2}
    \end{pmatrix}, \quad \begin{pmatrix}
        \frac{1}{2} & 0 \\ 0 & \frac{1}{2}
    \end{pmatrix}, \quad \begin{pmatrix}
        0 & 0 \\ 0 & 1
    \end{pmatrix}
    \label{eq:evolve}
\end{eqnarray}

Therefore, there is no locally measurable feature of $M$ that is quantum. In this simple toy model, using fermions as an example, we see how a locally classical system can indeed locally mediate quantum entanglement. However, one can wonder if the price to pay is that during the process, the classical system will witness some form of stochasticity as the pure to mixed and mixed to pure we observe in Eq. \ref{eq:evolve}. 

\subsection{Non-abelian anyons example}\label{subsec:nonabelian}

\begin{figure}[hb!]
    \centering
    \includegraphics[width=0.81\linewidth]{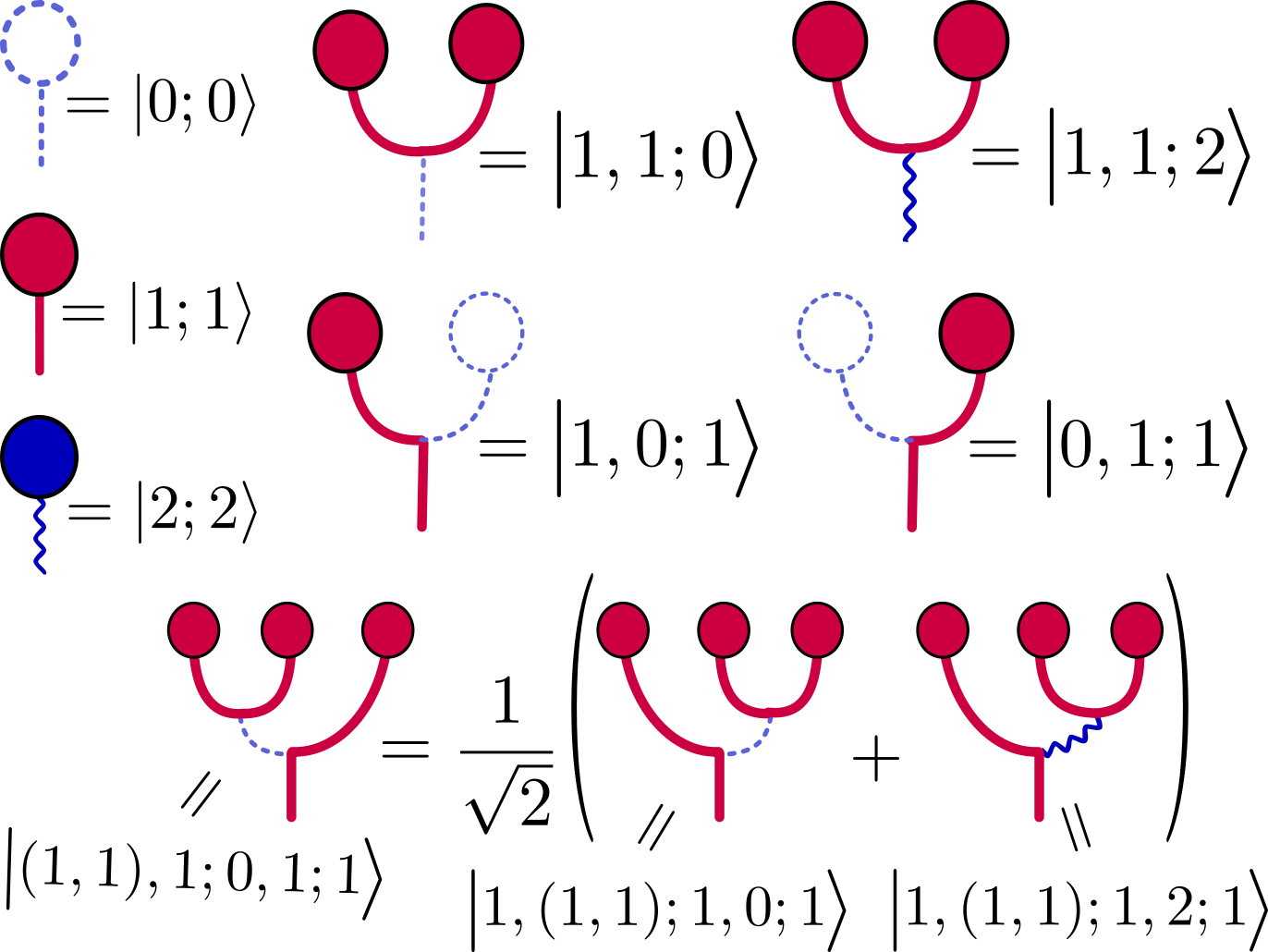}
    \caption{Diagrammatic and Dirac representations of Ising anyons}
    \label{fig:ising}
\end{figure}

\begin{figure}[hb!]
    \centering
    \includegraphics[width=\linewidth]{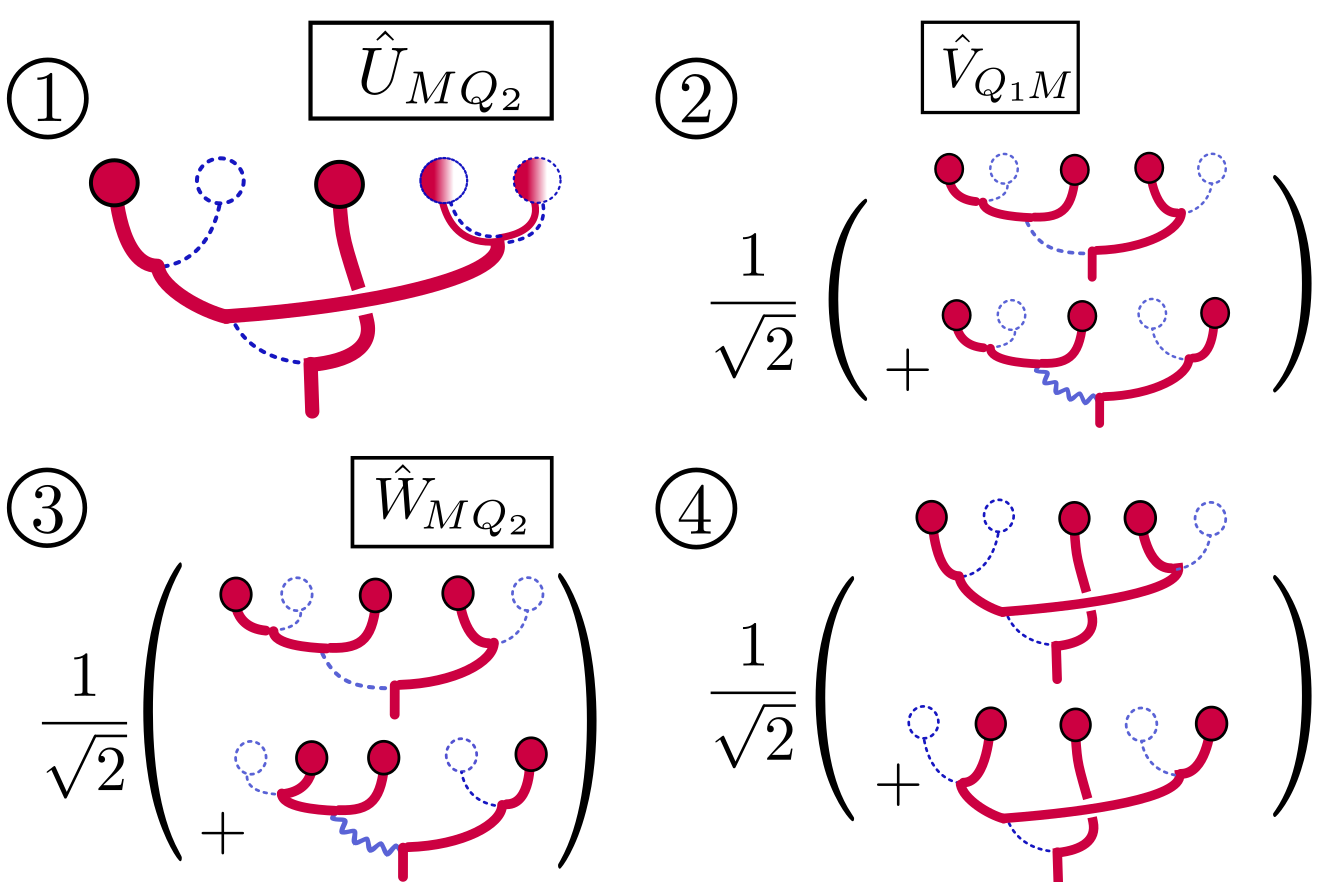}
    \caption{Diagram displaying the second toy model setup using the diagrammatic anyonic language in which is based on}
    \label{fig:anyons}
\end{figure}

In this example we show how we can have a locally classical system that can locally mediate entanglement without witnessing any form of stochasticity. We showcase this by using a constrained quantum system again. Specifically, we use a model based on the condensed matter system of non-Abelian Ising anyons \cite{nayak_non-abelian_2008, pachos_introduction_2012, xu_quantum_2022, simon_topological_2023}. Such particles are known to be connected with Majorana fermions \cite{campbell_majorana_2014, lutchyn_majorana_2018, kitaev_unpaired_2001}. We expect similar toy models can be found based on any non-Abelian anyon theory.  In Figure \ref{fig:ising}, we display the correspondence between the Dirac notation we use and the diagrammatic language used in the non-Abelian anyon literature \cite{pachos_introduction_2012, nayak_non-abelian_2008, bonderson_non-abelian_2007, simon_topological_2023}. Figure  \ref{fig:anyons} showcases the protocol we use diagrammatically.

The model we present is a direct translation of the non-Abelian anyon formalism widely used in the established fields of lattice gauge field theory \cite{pachos_introduction_2012} and topological quantum computing \cite{kitaev_fault-tolerant_2003, kitaev_anyons_2006, kitaev_unpaired_2001, freedman_topological_2003, freedman_topological_2001, simon_topological_2023, pachos_introduction_2012}. Let us introduce the basic notions with Dirac notation. 

Let us consider $5$ individual subsystems in a $ 1$D chain. The local space of states and observables for each of the $5$ individual subsystems is isomorphic to a maximally constrained qutrit. Therefore, the space is locally a classical trit. We denote the pure states of the $j$ individual subsystem as $\ket{x; x}_j$ where $x\in \{0,1,2\}$ and $j\in \{1,\dots,5\}$.  

The first particularity of these systems is that there is an explicit extra degree of freedom when systems are composed. That is due to the non-Abelian nature of the underlying anyonic theory. Concretely, in our case, an orthonormal basis of the states in the joint system of $2$ individual subsystems $j,k$ can be denoted as $\ket{x,y;z}_{j {k}}$ where $x,y,z\in \{0,1,2\}$ and they satisfy the rules expressed in Table \ref{tab:fusion}. One should interpret $x$ as the value of the individual system $j$ and $y$ as the one associated with $k$, and $z$ is a label that controls the degree of freedom of how the two subsystems are coupled together. Thus, there is an explicit degree of freedom in the coupling of the subsystems.   

\begin{table}[h]
    \centering
    \begin{tabular}{c|c c c}
        z & x=0 & x=1 & x=2 \\ \hline
        y=0 & 0 & 1 & 2 \\
        y=1 & 1 & 0 , 2 & 1\\
        y=2 & 2 & 1 & 0
    \end{tabular}
    \caption{Possible values of $z$ given $x$ and $y$ in $\ket{x,y;z}_{j {k}}$. }
    \label{tab:fusion}
\end{table}

Observe that both $\ket{1,1;0}_{j {k}}$ and $\ket{1,1;2}_{j {k}}$ are possible global states with both giving the value $1$ to both $j$ and $k$, however, they are orthogonal states. Thus, the states are globally distinguishable but not through local measurements. This is the intuition that tells us the couplings in the theory are non-local tomographic.

Notice also that Table \ref{tab:fusion} indicates that the state space of two joint individual subsystems has a dimension of $10$, whereas an individual one has a dimension of $3$. Both the individual subsystem and joint Hilbert spaces are superselected by ensuring that no superposition of states with different ''total charge'' $z$ is allowed. Thus, the $10$ dimensional joint space breaks down into blocks of dimension $3,4,3$ corresponding to the values of $z=0,1,2$ respectively. 

The last crucial property of our example is that the composition of individual subsystems is not associative. In other words, a state given by joining $j$ and $k$ first and then joining to $l$ is not the same as the one obtained by joining first $k$ and $l$ and then joining $j$ to the result. However, these two different compositions of $j,k,l$ are not identical but are isomorphic. It is a similar phenomenon to spin-network recombination basis \cite{de_pietri_spin_1997}. We denote for the first composition, $\ket{(x_0, x_1), x_2; z_{01}, x_2 ; g}_{(j {k}) l}$ where both $x_0, x_1, z_{01}$ and $z_{01}, x_2, g$ need to satisfy the relations established for triples $x,y,z$ in Table \ref{tab:fusion}. Similarly, we denote for the second $\ket{x_0, (x_1, x_2); x_{0}, z_{12} ; g}_{j ({k} {l})}$ where now both $x_1, x_2, z_{12}$ and $x_0, z_{12}, g$ need to satisfy the relations for triples $x,y,z$ in Table \ref{tab:fusion}. 

It is easy to check that both compositions yield the same dimension of $34$ and block structure of dimensions $10,14,10$ by imposing the SSR that only states with the same $g$ label are allowed to superpose. The isomorphism between the two compositions is easily expressed as the following unitary relations between the two orthonormal bases we described as $\ket{(x_0, x_1), x_2; z_{01}, x_2 ; g}_{(j {k}) l}= \ket{x_0, (x_1, x_2); x_{0}, z_{12}; g}_{j ({k} {l})}$ except for $x_0,x_1,x_2,g=1$, in which case we have
\begin{eqnarray}
    \ket{(1, 1), 1; 0, 1 ; 1}_{(j {k}) l}= \frac{1}{\sqrt{2}} \left(\ket{1, (1, 1); 1, 0; 1}_{j ({k} {l})}\right.\nonumber \\ \left. +\ket{1, (1, 1); 1, 2; 1}_{j ({k} {l})} \right) \label{eq:Fmove1} \\
    \ket{(1, 1), 1; 2, 1 ; 1}_{(j {k}) l} = \frac{1}{\sqrt{2}} \left(\ket{1, (1, 1); 1, 0; 1}_{j ({k} {l})}\right.\nonumber \\\left. - \ket{1, (1, 1); 1, 2; 1}_{j ({k} {l})} \right) \label{eq:Fmove2} 
\end{eqnarray}

These rules fix the structure we will be using. In our example, we consider five elemental subsystems. Two with charges $x_1,y_1$ compose $Q_1$, one with charge $m$ acts as the gravitational mediator $M$, and the other two with charges $x_2,y_2$ constitute $Q_2$. Notice that analogously to the fermionic case, $M$ is locally classical since it is maximally superselected and only the charge $m$ can be locally measured in $M$. Thus, $M$ is locally a classical trit. We use three different associations of the elemental subsystems in our entangling protocol. 

The first is to first associate $Q_1$ and $Q_2$, and then couple to $M$. We call it the matter-mediator partition. We express the states in such a partition $\ket{(x_1,y_1)),m,((x_2,y_2); z_1), m, (z_2 ; m, t; g}$. The global charges of $Q_1$ and $Q_2$ are $z_1,z_2$ and result from fusing $(x_1,y_1)$ and $(x_2,y_2)$ respectively. $t$ is the global charge of the matter sector $Q_1 Q_2$, thus $t$ is the fusion of $(z_1,z_2)$. Finally, $g$ is the global charge of the system, which results from fusing $t$ of the matter sector with $m$ in the mediator sector. 

The second partition we consider is the $Q_1 | M Q_2$ where we can interpret that the mediator is locally interacting with $Q_2$; then the result is coupled to the spatially-separated $Q_1$. We call it the right-hand-side partition. The states are expressed as $\ket{(x_1,y_1),(m, (x_2,y_2)); z_1, (m , z_2) ; z_1, h_2 ; g}$. Here, $h_2$ is the global charge of the sector $M Q_2$; thus, $h_2$ is the fusion of $(m,z_2)$. We need to know how the states with labels $t$ in the matter-mediator partition relate to states with labels $h_2$ in the right-hand side partition. 

Lastly, we consider the left-hand-side partition $Q_1 M | Q_2$ where $m$ interacts locally with $Q_1$. The states are expressed as  $\ket{((x_1,y_1),m), (x_2,y_2); (z_1, m) , z_2 ; h_1, z_2 ; g}$. $h_1$ is the global charge obtained by combining $(z_1,m)$. 

We need the relationship between the states in the three different partitions. We consider only states where $z_1=1=z_2=g=m$ and where $x_1,x_2,y_1,y_2 \in \{0,1\}$. Under these specifications, the possible values of $t,h_1.h_2$ are only $0$ and $2$. Thus, to relate a state in one partition with another we just need to specify a $3\times 3$ unitary matrix that follows from the underlying anyon theory. Concretely, we obtain that to change from left to right-hand-side partition, we use the matrix $P_{L\to R}$ which can be seen to come from Eq. \ref{eq:Fmove1} \& Eq. \ref{eq:Fmove2}. 

\begin{small}
    \begin{gather}
     \begin{aligned}
         &\ket{(x_1,y_1),(1, (x_2,y_2)); 1, (1 , 1) ;  1, h_2 ; 1}=   \nonumber \\   & \sum_{h_1} \left[ P_{L\to R}\right]_{h_1 h_2} \ket{((x_1,y_1),1), (x_2,y_2); (1, 1) , 1 ; h_1, 1 ; 1} 
     \end{aligned} \\  P_{L \to R} = \frac{1}{\sqrt{2}} \begin{pmatrix} 1 & 0 &  1 \\  0 & 0 & 0 \\1 & 0 & -1 \end{pmatrix}      
     \end{gather}
\end{small}

Similarly, to go from the matter-mediator partition to the right-hand side partition, we use the matrix $P_{C\to R}$. The complex phases come from the fractional statistics of the underlying Ising anyons theory we use \cite{pachos_introduction_2012, simon_topological_2023},
\begin{small}
    \begin{gather}
    \begin{aligned}
     & \ket{(x_1,y_1),(1, (x_2,y_2)); 1, (1 , 1) ;  1, h_2 ; 1}= \nonumber\\ &\sum_t \left[ P_{C\to R}\right]_{t h_2} \ket{(x_1,y_1)), 1, ((x_2,y_2); 1), 1 , (1 ; 1, t ; 1} 
     \end{aligned}\\ P_{C \to R} = \frac{1}{\sqrt{2}} \begin{pmatrix} 1 & 0 & 1 \\0 & 0 & 0\\ -i & 0 & i  \end{pmatrix} 
     \end{gather}
\end{small}

We can combine $P_{L\to R}$ and $P_{C\to R}$, along with their inverses, to convert states between two other partitions. For example $P_{L\to C}= \left(P_{C\to R}\right)^{-1} P_{L \to R} $. Let us notice that in all the states, we consider the charge $m$ of the mediator $M$ to be always $1$. Physically, this means that the local state of the mediator is always $1$ and remains unchanged by the interaction with the quantum matter. This will lead us to generate locally mediated entanglement without stochasticity in the locally classical mediator $M$. 

Before showcasing how entanglement in the matter sector can be generated, we first need to consider how we are treating the sectors $ Q_1$ and $ Q_2$ as quantum. We have fixed the global charges of these sectors to be $z_1=1=z_2$ in all states and observables we consider. The formalism and our simplifying restrictions allow us to have local superpositions of states in $Q_1$ and $Q_2$ of the form $\alpha \ket{1,0;1}+\beta \ket{0,1;1}$. We can consider these as an encoding of a qubit state $\alpha \ket{0} +\beta \ket{1}$ in the anyonic-inspired system. However, we need to specify the coupling degree of freedom $t$ between $Q_1$ and $Q_2$ that corresponds to our encoding of a 2-qubit system. We choose $t=0$ as the encoding for our two-qubit system. The encoding of the two-qubit computational states to the anyonic-inspired toy model is specified in Table \ref{tab:encoding}. 

\begin{table}[ht]
    \centering
    \begin{tabular}{c|c}
       2-qubit state  &   4-Ising state \\ \hline 
       $\ket{0}\otimes \ket{0}$   & $\ket{(1,0),(1,0);1,1;0}$ \\
        $\ket{0}\otimes \ket{1}$   & $\ket{(1,0),(0,1);1,1;0}$ \\
         $\ket{1}\otimes \ket{0}$   & $\ket{(0,1),(1,0);1,1;0}$ \\
          $\ket{1}\otimes \ket{1}$   & $\ket{(0,1),(0,1);1,1;0}$ \\
    \end{tabular}
    \caption{Encoding of the two-qubit states on the 4 Ising anyon system $Q_1 Q_2$.}
    \label{tab:encoding}
\end{table}

To discuss whether there is entanglement in an encoded state, we specify the system-local observables of our quantum systems $Q_1$ and $Q_2$. First, let us consider the $X_1$ and $Z_1$ equivalent observables in $Q_1$
\begin{align}
    \hat{X}_1 = \ketbra{(1,0);1}{(0,1); 1} + \ketbra{(0,1);1}{(1,0);1} \label{eq:X}\\ 
    \hat{Z}_1 = \ketbra{(1,0);1}{(1,0); 1} - \ketbra{(0,1);1}{(0,1);1}
\end{align}

Their representation in the composite system $Q_1 Q_2$ as system-local observables is 
\begin{small}
    \begin{align}
    \hat{X}_1  = \sum_{x_2,t} \ketbra{(1,0),(x_2,1-x_2);1,1;t}{(0,1),(x_2, 1-x_2); 1,1;t}  \nonumber \\ +\sum_{x_2,t} \ketbra{(0,1),(x_2,1-x_2);1,1;t}{(1,0),(x_2, 1-x_2); 1,1;t} \label{eq:embedX} \\ 
    \hat{Z}_1 = \sum_{x_2,t}  \ketbra{(1,0),(x_2,1-x_2);1,1;t}{(1,0),(x_2, 1-x_2); 1,1;t} \nonumber \\ -\sum_{x_2,t} \ketbra{(0,1),(x_2,1-x_2);1,1;t}{(0,1),(x_2, 1-x_2); 1,1;t} \label{eq:embedZ}
\end{align}
\end{small}

And completely analogously to the system-local observables in $Q_2$. The sum over $x_2$ ranges from 0 to 1, while the sum over $t$ takes the values 0 and 2. These observables do not commute as in the usual qubit case. Moreover, the charges associated with sector $Q_2$ are left invariant, as expected. Identifying system-local observables leads to the unique characterisation of the partial tracing procedure \cite{krumm_quantum_2021, vidal_quantum_2021}. Both the partial tracing operation and the system-local observables notion come directly from the anyonic formalism \cite{bonderson_non-abelian_2007, simon_topological_2023}.  The partial trace over $Q_2$ of a general operator in $Q_1 Q_2$ is given by:
\begin{small}
\begin{flalign}
    \Tr_{Q_2} & \left(\ketbra{(x_1,y_1),(x_2,y_2);z_1,z_2;g}{(\tilde{x}_1,\tilde{y}_1),(\tilde{x}_2,\tilde{y}_2);\tilde{z}_1,\tilde{z}_2;g}\right) \nonumber \\  &=\delta_{x_2 \tilde{x}_2} \delta_{y_2 \tilde{y}_2} \delta_{z_2 \tilde{z}_2} \delta_{z_1 \tilde{z}_1} \ketbra{(x_1,y_1);z_1}{(\tilde{x}_1,\tilde{y}_1);\tilde{z}_1} \label{eq:ptQ} 
\end{flalign}
\end{small}

Crucially, the partial trace over the mediator $M$ in the composite system $Q_1 M Q_2$ is expressed concisely in the matter-mediator partition as 
\begin{small}
\begin{gather}
    \ket{\psi}=\ket{(x_1,y_1)), m , ((x_2,y_2);z_1),m, (z_2;m,t;g}\nonumber \\ \bra{\eta}=\bra{(\tilde{x}_1,\tilde{y}_1)), \tilde{m}, ((\tilde{x}_2,\tilde{y}_2);\tilde{z}_1),\tilde{m},(\tilde{z}_2;\tilde{m},\tilde{t};g}  \nonumber \\ \begin{aligned} & \Tr_M\left(\ketbra{\psi}{\eta}\right)= \\ &\delta_{m \tilde{m}} \delta_{t \tilde{t}} \ketbra{(x_1,y_1), (x_2,y_2);z_1,z_2;t}{(\tilde{x}_1,\tilde{y}_1), (\tilde{x}_2,\tilde{y}_2);\tilde{z}_1,\tilde{z}_2;\tilde{t}} \label{eq:ptm} \end{aligned}
\end{gather}
\end{small}

Since, in essence, we are working with constrained quantum systems that do not display a tensor product structure per se, we need to specify how to measure entanglement. We only need to focus on pure states in $Q_1 Q_2$. Following the literature \cite{nielsen_quantum_2000, vidal_entanglement_2024}, any bipartite pure state that is not uncorrelated is an entangled state. A bipartite pure state $\ket{\psi}_{Q_1 Q_2}$ is uncorrelated if and only if it satisfies Equation \ref{eq:uncorr}. This notion of an uncorrelated pure state follows the requirement that the expectation values of the products of system-local observables be a product of the local expectation values. We use it in all three examples.  

The last necessary ingredient in the formalism is identifying unitaries in $Q_1 M Q_2$ that are system-local in $Q_1 M$ or $M Q_2$. To have a consistent theory, unitaries must preserve the global charge of the system they act on. Embedding system-local unitaries into composite systems follows the same procedure as in Eq. \ref{eq:embedX} \& Eq. \ref{eq:embedZ}. Consider a unitary $\hat{U}$ of $Q_1 M$
\begin{gather}
    \hat{U}= \sum_{\vec{v}, \vec{w}, h_1 } u_{\vec{w} h_1}^{\vec{v} h_1} \ketbra{\vec{v};h_1}{\vec{w};h_1}  \label{eq:localU}
\end{gather}

 where we use the shortcut notation $\ket{\vec{v};h_1}=\ket{(x_1,y_1),m; z_1,m; h_1}$ and  $\bra{\vec{w};h_1}=\bra{(\tilde{x}_1,\tilde{y}_1),\tilde{m}; \tilde{z}_1,\tilde{m}; h_1}$, and the sums correspond to all the allowed combinations of values of the charges, according to Table \ref{tab:fusion}. Then, the embedding of this system-local unitary to the composite system $Q_1 M Q_2$ is given by 
\begin{small}
\begin{gather}
    \scalebox{0.93}{$\hat{U}= \displaystyle \sum_{\substack{\vec{v}, \vec{w}, h_1, x_2 \\ y_2,z_2, g} } u_{\vec{w} h_1}^{\vec{v} h_1} \ketbra{\vec{v},(x_2,y_2);h_1,z_2;g}{\vec{w},(x_2,y_2);h_1,z_2;g}$}   \label{eq:embeddedU}
\end{gather}
\end{small}

where we use the shortcut $\ket{\vec{v},(x_2,y_2);h_1,z_2;g}=\ket{((x_1,y_1),m),(x_2,y_2);(z_1,m),z_2;h_1,z_2;g}$. The expression sums over all the possible combinations of values of the labels allowed by Table \ref{tab:fusion}. The unitary leaves any charge associated with $Q_2$ completely unchanged, exactly as expected for a system-local unitary in $Q_1 M$. The system-local unitaries in $M Q_2$ have the equivalent embedding procedure.    

We are ready to present the protocol that sends a pure separable state in $Q_1Q_2$ to a pure entangled state through system-local interactions of $Q_1M$ and $MQ_2$. It is diagrammatically displayed in Figure \ref{fig:anyons}. 
The initial separable state is the encoding of the $\ket{0}\otimes \ket{+}$ 2-qubit state. The mediator is initialised at $m=1$. Therefore, the initial pure state we consider is 
\begin{flalign}
    \ket{\psi_0}=\frac{1}{\sqrt{2}} & \left( \ket{(1,0)),1,((1,0);1),1,(1;1,0;1}+\right. \nonumber \\ & \left. \ket{(1,0)),1,((0,1);1),1,(1;1,0;1}\right)
\end{flalign}

Using the partial tracing procedures we introduced and the notion of uncorrelated states in $Q_1 Q_2$, it is straightforward to check that 
\begin{eqnarray}
    \Tr_{M}\left(\ketbra{\psi_0}\right) = \frac{1}{\sqrt{2}} \left(\ket{(1,0),(1,0);1,1;0}\right.\nonumber \\ \left. +\ket{(0,1),(0,1);1,1;0}\right)
\end{eqnarray} 
and that such a state is a bipartite pure uncorrelated state in $Q_1 Q_2$. 

In our protocol, we act with the following sequence of local unitaries: $\hat{U}_{M Q_2}, \hat{V}_{Q_1 M}, \hat{W}_{M Q_2}$ local in $M Q_2$, $Q_1 M$ and $M Q_2$, respectively. The local unitaries we choose are such that they leave all the states invariant except for:
\begin{align}
    & \hat{U}_{M Q_2} \ket{1,(1,0);1,1;2}= i \ket{1,(1,0);1,1;2}  \\
     & \hat{U}_{M Q_2} \ket{1,(0,1);1,1;2}= -i \ket{1,(0,1);1,1;2} \\
     & \hat{V}_{Q_1 M} \ket{(1,0),1;1,1;2}=  \ket{(0,1),1;1,1;2} \\
      & \hat{V}_{Q_1 M} \ket{(0,1),1;1,1;2}=  \ket{(1,0),1;1,1;2} \\
     & \hat{W}_{M Q_2} \ket{1,(1,0);1,1;2}=  -i \ket{1,(1,0);1,1;2} \\
     & \hat{W}_{M Q_2} \ket{1,(0,1);1,1;2}=  +i \ket{1,(0,1);1,1;2} 
\end{align}

These specifications correspond to $\hat{U}_{M Q_2}$ being a controlled $i Z$ gate with the control being $h_2$ and the target our embedded qubit in $Q_2$ as $\ket{0}=\ket{(1,0);1}$ and $\ket{1}=\ket{(0,1);1}$. The presence of $m=1$ just acts to access the quantum degree of freedom of the coupling between $M$ and $Q_2$. We have a superposition in such a degree of freedom due to the non-Abelian allowed extra states in the composition and the non-associativity of the partitions. 

This specific mechanism allows a locally classical system coupled to quantum systems and entanglement to be generated through local interactions. Similarly, $\hat{V}_{Q_1 M}$ is a controlled $X$ gate in $Q_1 M$ in the same sense. $\hat{W}_{ M Q_2}$ is a controlled $-i Z$, but in $M Q_2$. After this insight, let us show the computation step by step. First $\hat{U}_{M Q_2} \ket{\psi_0}$ is equal to
\begin{align}
     \hat{U}_{M Q_2} \frac{P_{C\to R}}{\sqrt{2}} & \left(\ket{(1,0)),1,((1,0);1),1,(1;1,0;1}\right. \nonumber \\ &\left. + \ket{(1,0)),1,((0,1);1),1,(1;1,0;1}\right) = \nonumber \\ \end{align}
which is
\begin{align}
\hat{U}_{M Q_2} \frac{1}{2} & \left(\ket{(1,0),(1,(1,0));1,(1,1);1,0;1}\right. \nonumber \\   &-i \ket{(1,0),(1,(1,0));1,(1,1);1,2;1} \nonumber \\   &+ \ket{(1,0),(1,(0,1));1,(1,1);1,0;1} \nonumber \\  &\left. -i \ket{(1,0),(1,(0,1));1,(1,1);1,2;1} \right) =\nonumber\\ 
      =\frac{1}{2} &\left(\ket{(1,0),(1,(1,0));1,(1,1);1,0;1}\right. \nonumber \\  &+ \ket{(1,0),(1,(1,0));1,(1,1);1,2;1} \nonumber \\   &+ \ket{(1,0),(1,(0,1));1,(1,1);1,0;1} \nonumber \\  &\left. - \ket{(1,0),(1,(0,1));1,(1,1);1,2;1} \right)
\end{align}

We have first converted the state from the matter-mediator partition to the right-hand side partition. Then, we have applied the definition of the system-local unitary $\hat{M Q_2}$. We express this state in the left-hand side partition. Using $P_{R \to L}$, we obtain $\hat{U}_{M Q_2} \ket{\psi_0}$ is 
\begin{align}
     \frac{1}{\sqrt{2}} & \left(\ket{((1,0),1),(1,0);(1,1),1;0,1;1}+ \right. \nonumber \\ & \left. +\ket{((1,0),1),(0,1);(1,1),1;2,1;1}\right)
\end{align}
We apply the local unitary $\hat{V}_{Q_1 M}$, thus getting $\hat{V}_{Q_1 M} \hat{U}_{M Q_2} \ket{\psi_0}$ equal to
\begin{align}
     \frac{1}{\sqrt{2}} &\left(\ket{((1,0),1),(1,0);(1,1),1;0,1;1}+ \right. \nonumber \\ & \left. +\ket{((0,1),1),(0,1);(1,1),1;2,1;1}\right)
\end{align}
which converted to the right-hand side partition is expressed as:
\begin{align}
    \frac{1}{2} & \left(\ket{(1,0),(1,(1,0));1,(1,1);1,0;1}\right. \nonumber \\  &+ \ket{(1,0),(1,(1,0));1,(1,1);1,2;1} \nonumber \\  & + \ket{(0,1),(1,(0,1));1,(1,1);1,0;1} \nonumber \\ & \left. - \ket{(0,1),(1,(0,1));1,(1,1);1,2;1} \right)
\end{align}
The last interaction is given by $\hat{W}_{M Q_2}$, obtaining the result of $\hat{W}_{M Q_2} \hat{V}_{Q_1 M} \hat{U}_{M Q_2} \ket{\psi_0}=$
\begin{align}
     =\frac{1}{2} & \left(\ket{(1,0),(1,(1,0));1,(1,1);1,0;1}+ \right. \nonumber \\  &-i \ket{(1,0),(1,(1,0));1,(1,1);1,2;1} \nonumber \\  & + \ket{(0,1),(1,(0,1));1,(1,1);1,0;1} \nonumber \\ &\left.-i \ket{(0,1),(1,(0,1));1,(1,1);1,2;1} \right)
\end{align}
which converting back to the matter-mediator partition the final state $\ket{\psi_f}$ ends up being
\begin{align}
     \ket{\psi_f}=\frac{1}{\sqrt{2}} & \left(\ket{(1,0)),1,((1,0);1),1,(1;1,0;1}+ \right. \nonumber \\  &\left. +\ket{(0,1)),1,((0,1);1),1,(1;1,0;1} \right)
\end{align}
The final state expression is very suggestive. By applying the partial trace of the locally-classical mediator system $M$ presented in Eq. \ref{eq:ptm}, we obtain the pure state $\ket{\eta}_{Q_1 Q_2}$ in the matter sector $Q_1 Q_2$ with 
\begin{align}
     \ket{\eta}_{Q_1 Q_2}=\frac{1}{\sqrt{2}} & \left(\ket{(1,0),(1,0);1,1;0}+ \right. \nonumber \\  & \left. +\ket{(0,1),1,(0,1);1,1;0} \right)
\end{align}
The state $\ket{\eta}_{Q_1 Q_2}$ is the encoding of the two-qubit state $\frac{1}{\sqrt{2}} \left(\ket{0}\otimes \ket{0} + \ket{1}\otimes \ket{1}\right)$ in $Q_1 Q_2$. The state $\ket{\eta}_{Q_1 Q_2}$ is entangled as the state it encodes. $\ket{\eta}_{Q_1 Q_2}$ is entangled if and only if it is not an uncorrelated state. Thus, if we can find a pair of system-local observables $\hat{A}_{Q_1}, \hat{B}_{Q_2}$ for which Eq. \ref{eq:uncorr} does not hold, then the state is entangled. 

By applying the anyonic partial tracing presented in Eq. \ref{eq:ptQ} we obtain that the reduced states in $Q_1$ and $Q_2$ are $\rho_{Q_1}=\frac{1}{2}\left(\ketbra{1,0;1}+\ketbra{0,1;1}\right)=\rho_{Q_2}$. The local quantum observables $\hat{X}_1$ and $\hat{X}_2$ presented in Eq. \ref{eq:X} do not satisfy the equality of Eq. \ref{eq:uncorr}. Indeed, the expected value of $\hat{X}_1 \cdot \hat{X}_2$ is $1$. Meanwhile, the product of local expected values of $\hat{X}_1$ and $\hat{X}_2$ is $0$.

We have demonstrated that a locally classical mediator can generate entanglement between two quantum systems through system-local interactions. Further, the mediator's local state remains pure throughout the interaction. Thus, no stochasticity can be associated with the mediator's local behaviour. The key is that we have exploited the possibility of coupling the mediator with the matter in such a way that the local properties of the mediator do not need to vary to unlock some extra degree of freedom on the boundary of the $M$ and $Q_j$ sectors. We constructed local interactions where the quantum properties of the hidden boundary degree of freedom were hedged to generate entanglement between $Q_1$ and $Q_2$ without them directly interacting. 

A way to interpret this phenomenon is that the mediator acts like a locally classical key to a quantum degree of freedom on the boundary. One can imagine the charge of the mediator $m$ being this key and is needed to be carried to interact with the relevant $Q_j$ to access the quantum degree of the boundary between $M$ and $Q_j$ and generate interactions controlled on the boundary quantum degree of freedom. Simultaneously, the boundary quantum degree of freedom is not accessible without the presence of any of the $Q_j$. Thus, the mediator is locally classical.  

We believe this example can be extremely relevant in a realistic study of the gravitational-matter interaction. If one understands anyonic systems as constrained quantum systems, one can think that the properties of the non-Abelian anyons that we have used arise due to the non-Abelian nature of the constraint in the quantum system. This observation leads us to the grounded speculation that in quantum field theories with strong/particular non-Abelian gauges or constraints, the structures we observe in this toy model are expected to appear. Thus, the phenomena observed can also be expected. Despite being non-perturbatively renormalisable, linear quantum gravity (LQG) is a gauged non-Abelian quantum field theory \cite{flanagan_basics_2005, christodoulou_locally_2023, maggiore_3_2007, maggiore_5_2007}. Similarly, other proposed theories of quantum gravity, such as loop quantum gravity and string theory, have direct relations and analogues with non-Abelian quantum field theories such as lattice quantum chromodynamics \cite{kuti_lattice_2006, delcamp_fusion_2017}.    

Nevertheless, our toy theory has three critical shortcomings. The first is the physical interpretation of the interactions. Nonetheless, in a complete theory with these properties, we could find reasonable physical interactions that would generate entanglement by exploiting the same method and characteristics. 

The second point is more challenging to tackle. If one looks carefully, one sees that the matter sector $Q_1 Q_2$ does not correspond to the space of two qubits; instead, it is larger. One could fine-tune the toy theory so $Q_1$ and $Q_2$ are two-dimensional. However, we believe it would be more challenging to consistently have a model where the dimensionality of $Q_1 Q_2$ is four. The explanation is that in our toy model, the degree of freedom of the global charge of $Q_1 Q_2$ (label $t$) is directly related to the degree of freedom in the boundaries of $M Q_j$ (label $h_j$). Thus, we reason that if one limits the values of $t$ to just $0$, then one would also restrict the values of $h_j$ to only $0$. In that case, our protocol would fail since we heavily rely on the property of $h_j$ being able to take both values $0$ and $2$. After all, we do control operations with them.       

The third point of contention is indirectly linked to the second. In our toy example, we have a single charge for the mediator $M$; however, if one were to consider more charges constituting the mediator, that would fail to be locally classical, as local non-commuting observables would appear.

Points two and three are interesting and relevant for future studies of non-locally tomographic couplings and their consistency. Nevertheless, we aim to provide toy examples based on existing structures and systems to raise awareness of the role of local tomography in the BMV arguments. Our work aims not to provide a complete theory that models gravitational interaction with non-locally tomographic couplings. Ultimately, our goal is to inspire future research avenues and raise awareness of this realm of possibilities.

\subsection{Classical and anticlassical bits} 
\label{subsec:bitanti} 

In this section, we present our third and final example of non-locally tomographic coupling that can generate entanglement with a locally classical mediator and system-local interaction. This example aims to cover some of the shortcomings of the other two. In particular, this example tackles the criticisms of the mediator system's small dimensionality and the impossibility of considering it constituted of several particles/subsystems within the toy theories. 

This example is based on the fully fledged toy theory of classical and anti-classical bits \cite{chiribella_bell_2024}. We want mediators that, irrespective of the size and number of degrees of freedom, behave classically and only act quantumly when coupled to matter. Let us briefly present the toy theory of classical-anticlassical bits. 

The elemental systems of the toy theory are bits $B$ and anti-bits $A$. Both are associated with a 2-dimensional Hilbert space $\pazocal{H}_A \simeq \mathbb{C}^2 \simeq \pazocal{H}_B$. A system $S$ consisting of only $m$ anti-bits or only $n$ bits is denoted as $(m,0)$ and $(0,n)$. In the case $(m,0)$, the only allowed pure states are of the form $\ket{\psi}_S= \ket{i_1}\otimes \dots\otimes \ket{i_m} \in \pazocal{H}_{A_1}\otimes \dots \otimes \pazocal{H}_{A_m} $ where $i_k \in \{0,1\}$ denote the states in the anti-bits computational basis. Similarly, for a system $(0,n)$, the pure states are of the form $\ket{\psi}_S= \ket{i_1}\otimes \dots\otimes \ket{i_m} \in \pazocal{H}_{B_1}\otimes \dots \otimes \pazocal{H}_{B_n} $ where $i_k \in \{0,1\}$ denote the bits computational basis.   

Notice that no state superposition is allowed in these systems. Therefore, they behave like a classical system of $n$ and $m$ bits, respectively. In this toy theory, bits couple to bits in a completely classical way. However, systems that contain both bits and anti-bits exhibit non-classical features.  Bits couple to anti-bits in a non-classical way. Consider a system $(m,n)$ of $n$ bits and $m$ anti-bits. We fix $n \geq m$. The pure states in such a system are of the form 
\begin{eqnarray}
    \sum_{\vec{i}} \lambda_{\vec{i}} \ket{\vec{i},\sigma,\vec{q}} \otimes \ket{j_{m+1}} \otimes \dots \otimes \ket{j_{n}}
\end{eqnarray}
where $\ket{j_{m+1}} \otimes \dots \otimes \ket{j_{n}}$ is an $n-m$ bit computational basis state with $j_k\in \{0,1\}$, all the vectors have length $m$, $\lambda_{\vec{i}} \in \mathbb{C}$, $\sum_{\vec{i}} |\lambda_{\vec{i}}|^2 =1$, and where  $\ket{\vec{i},\sigma,\vec{q}}$ denotes a state in $\pazocal{H}_{A_1}\otimes \dots \otimes \pazocal{H}_{A_m} \otimes \pazocal{H}_{B_1}\otimes \dots \otimes \pazocal{H}_{B_m} \simeq (\mathbb{C}^2)^{\otimes 2 m} $ given by $\ket{\vec{i},\sigma,\vec{q}}=$
\begin{flalign}
     \scalebox{0.95}{$\displaystyle \ket{i_1}\otimes \dots \otimes \ket{i_m} \otimes \ket{i_{\sigma(1)} \oplus q_1} \otimes \dots \otimes \ket{i_{\sigma(m)} \oplus q_m}$}
\end{flalign}
with $\sigma \in S_m$ a permutation of $m$ elements, $i_k, q_k\in \{0,1\}$ for all $k$ from $1$ to $m$, and the $\oplus$ denotes addition modulo $2$. $\ket{\vec{i},\sigma,\vec{q}}$ is the specification of the tensor product of computational basis states for the $m$ bits and anti-bits under a specific matching given by a permutation $\sigma \in S_n$ that assigns a bit with some anti-bit and the vector $\vec{q}$ that specifies the allowed compositions of bits and anti-bits. Notice that such composition is non-classical since it allows certain superpositions of computational basis states. We can superpose states with the same $\vec{q}$ and $\sigma$.  

This toy theory has the advantage of being based on the usual tensor product. Thus, the tensor product structure provides the typical notion of system locality. This construction enables the standard partial tracing procedure for tensor product systems to be applied. This fact is very useful since we can easily characterise whether a pure state is entangled. Nevertheless, since the allowed states of bit and anti-bit composites are strictly larger than the classical composition, the theory is non-locally tomographic. We proceed to demonstrate how we can utilise this theory, which incorporates classical theory, to generate entanglement through local gravitational interactions, where the gravitational mediator is represented by classical bits only.   

\begin{figure}[ht]
    \centering
    \includegraphics[width=\linewidth]{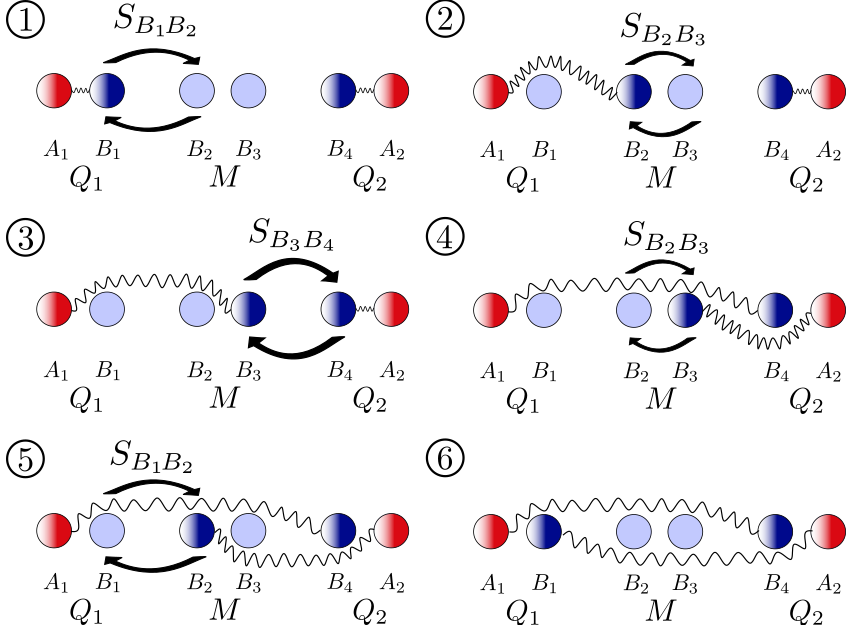}
    \caption{Diagram displaying how bits and anti-bits can generate entanglement through local interaction with a mediator of two classical bits. Bits in blue, anti-bits in red. The half-filled linked bit-anti-bot pairs denote a superposition in the quantum system they form.}
    \label{fig:bits}
\end{figure}

To fix a reasonable dimension and focus on our example, let us fix a system $(2,4)$ consisting of two anti-bits and four classical bits. We are interested in a scenario where a pair of a bit and an anti-bit encodes a qubit, two bits represent the gravitational mediation, and the remaining bit and anti-bit pair encodes a second qubit. Therefore, the labelling and ordering of the bits and anti-bits we use is $A_1 B_1 B_2 B_3 B_4 A_2$. $(A_1 B_1)$ encode one qubit, $(B_4 A_2)$ the other, and $(B_2, B_3)$ encode the gravitational mediator. See Figure \ref{fig:bits} for a diagrammatic representation of our proposed example. It is similar to the fermionic case in Figure \ref{fig:fermion}, but with this theory, we can overcome the problem where the dimension of the mediator was limited by the fermionic algebra becoming locally quantum when more than one fermionic mode is considered.    

We use the minimal choice of dimension to showcase that, using this theory, there is no limit to the number of classical degrees of freedom that the gravitational sector can have. Extending our example to one where the number of classical bits associated with the gravitational sector is as large as desired is straightforward.    

The only required physical transformation in the protocol is the SWAP gate between two bits. For simplicity, we drop the tensor product symbol from now on. We denote the swapping gate of bits $B_1$ and $B_2$ as $S_{B_1 B_2} \ket{x}_{B_1}  \ket{y}_{B_2} = \ket{y}_{B_1}  \ket{x}_{B_2}$. Such an operation is allowed in the theory, and it is a direct classical transformation since it acts locally only in the classical bits sector of the theory. Moreover, it is system-local in the two bits being swapped. Let us fix the encoding of a qubit in each bit anti-bit pair we use. We encode a general qubit state $\ket{\psi}=\alpha \ket{0} + \beta \ket{1}$ in a bit anti-bit pair $A B$ as $\ket{\tilde{\psi}}=\alpha \ket{0}_A \ket{0}_B + \beta \ket{1}_A \ket{1}_{B}$. This corresponds to choosing the composition of the bit and anti-bit pair state space given by $q=0$ for their pairing. However, notice that the states corresponding to choosing $q=1$ are also allowed. These correspond to states of the form $\alpha_1 \ket{0}_A \ket{1}_B + \beta_1 \ket{1}_A \ket{0}_{B}$. One way to interpret this state space is that a bit anti-bit pair corresponds to having a qubit and a bit. The bit chooses in which sector of the state space the state is located, either $q=0$ or $q=1$. Even though we choose to encode our qubit state in the $q=0$ sector, we need to remember that the $q=1$ states are still allowed and part of the state space. To avoid being overwhelmed with notation, we drop the labels of bit and anti-bit and consider them fixed in the order we have established unless explicitly stated otherwise.   
 
The protocol we consider starts with the pure state $\ket{\psi_0}$ that encodes two $\ket{+}$ states in the qubits, and the gravitational mediator is in the zero computational state, obtaining
\begin{small}
\begin{gather}
    \scalebox{0.92}{$\displaystyle \frac{\left(\ket{0}_{A_1}  \ket{0}_{B_1} + \ket{1}_{A_1}  \ket{1}_{B_1} \right)}{\sqrt{2}}  \ket{0}_{B_2}  \ket{0}_{B_3}  \frac{\left(\ket{0}_{B_4} \ket{0}_{A_2} + \ket{1}_{B_4} \ket{1}_{A_2}\right)}{\sqrt{2}}$} \nonumber \\ \equiv \frac{\left(\ket{0 0} + \ket{1 1}  \right)}{\sqrt{2}}  \ket{0 0} \frac{\left(\ket{0 0} + \ket{1 1} \right)}{\sqrt{2}} 
\end{gather}
\end{small}
The sequence of system local interactions we apply is this sequence of classical bit swap gates $S_{B_1 B_2}, S_{B_2 B_3}, S_{B_3 B_4}, S_{B_2 B_3}, S_{B_1 B_2}$. Intuitively, this corresponds to bringing the bit that constitutes the first qubit to the second and vice versa by swapping them with the mediating bits that constitute the gravitational mediator. Indeed, it can be verified that after this sequence of operations, we obtain
\begin{small}
    \begin{align}
    & \medspace S_{B_1 B_2} S_{B_2 B_3} S_{B_3 B_4} S_{B_2 B_3} S_{B_1 B_2} \ket{\psi_0}\!=\nonumber \\  &=\!S_{B_1 B_2} S_{B_2 B_3} S_{B_3 B_4} S_{B_2 B_3} \frac{1}{2} \left(\ket{0 0 0 0}\! +\!\ket{1 0 1 0} \right)\! \left(\ket{0 0}\! +\!\ket{1 1} \right) \nonumber \\ &=\!S_{B_1 B_2} S_{B_2 B_3} S_{B_3 B_4}  \frac{1}{2} \left(\ket{0 0 0 0 }\! +\!\ket{1 0 0 1}\! \right)\! \left(\ket{0 0}\!+\!\ket{1 1} \right) \nonumber \\  &=\!S_{B_1 B_2} S_{B_2 B_3} \frac{1}{2} \left(\ket{0 0 0 0 0 0}\! +\!\ket{1 0 0 0 1 0}\! +\!\ket{0 0 0 1  0 1}\! +\!\ket{1 0 0 1 1 1} \right)  \nonumber \\  &=\!S_{B_1 B_2} \frac{1}{2} \left(\ket{0 0 0 0 0 0}\! +\!\ket{1 0 0 0 1 0}\! +\!\ket{0 0 1 0  0 1}\! +\!\ket{1 0 1 0 1 1} \right)\nonumber \\ &=\!\frac{1}{2} \left(\ket{0 0 0 0 0 0}\! +\!\ket{1 0 0 0 1 0}\! +\!\ket{0 1 0 0 0 1}\! +\!\ket{1 1 0 0 1 1} \right) 
\end{align}
\end{small}
At the end of the protocol, the two bits that belong to the gravitational mediator are in the zero computational state, separable from the two-bit anti-bit pairs. Indeed, by taking the partial trace over the bits $B_2 B_3$, we see that the state of the quantum sector of the protocol is pure, and its expression is 
\begin{eqnarray}
\ket{\psi}_Q = \frac{1}{2} \left(\ket{0 0  0 0} + \ket{1 0  1 0} + \ket{0 1  0 1} + \ket{1 1  1 1} \right)    \label{eq:q}
\end{eqnarray}
Such a state encodes a pure quantum bipartite state. However, each of the bit-anti-bit pairs can be seen as a qubit and a bit, where the bit indicates the superselection sector to which the qubit belongs, either $q=1$ or $q=0$. Taking  the partial trace over $A_1 B_1$, we obtain the local state of the bit anti-bit pair $B_4 A_2$ being 
\begin{eqnarray}
\rho_{Q_2} = \frac{1}{4} \left(\ketbra{0 0} + \ketbra{0 1} + \ketbra{1 0} + \ketbra{1 1} \right)    
\end{eqnarray}
which is a maximally mixed state in the $B_4 A_2$ system. Due to the symmetry of the state in Equation \ref{eq:q}, we obtain $\rho_{Q_2}=\rho_{Q_1}$. Thus, both marginals of the bipartite quantum pure state are maximally mixed. Therefore, we expect the pure state in Equation \ref{eq:q} to be entangled. To be sure, let us provide two local observables of $Q_1$ and $Q_2$, $\hat{X}_1, \hat{X}_2$ such that the uncorrelated state equation is not satisfied 
\begin{eqnarray}
 \Tr(\hat{X}_1   \rho_{Q_1}) \Tr( \hat{X}_2 \rho_{Q_2}) =  \Tr(  \hat{X}_1 \otimes \hat{X}_2 \ketbra{\psi}_Q) 
\end{eqnarray}

The local operators $\hat{X}_1=\left(\ketbra{00}{11}+\ketbra{11}{00}\right) \otimes \mathbb{I}$ and $\hat{X}_2=  \mathbb{I} \otimes \left(\ketbra{00}{11}+\ketbra{11}{00}\right)$ violate the equality above. The left-hand side gives a value of $0$, and the right-hand side is $\frac{1}{2}$. The result showcases that the final state of the matter sector is entangled. Therefore, we demonstrated how, starting from a separable state and having only local interactions between the matter sector and the gravitational mediator, we can generate entanglement. Moreover, we have seen that in this toy model, there is no restriction on the number of degrees of freedom that the locally classical gravitational mediator can have.  

\section{Discussion}\label{sec:discuss}

In this paper, we showcase how spacetime can mediate entanglement while violating the usual assumption of local tomography. This shows how local tomography is not necessary for the GWT or to model the BMV effect. The models of spacetime we consider do not couple to matter satisfying local tomography. Despite being completely classical locally, by definition, a non-locally tomographic coupling of spacetime and matter classifies spacetime as non-classical \cite{galley_any_2023}. 

We present three toy theories that exemplify how a non-locally tomographic coupling of a gravitational mediator with quantum matter can lead to entanglement generation through system-local interaction. We show that a locally classical gravitational mediator can generate entanglement through system-local interactions with quantum matter if the coupling of the two distinct sectors does not satisfy local tomography. Each of the three examples we have provided sheds light on the possibilities and shortcomings of the system-local entanglement generation. 

In the fermionic example, we demonstrate how a simple superselection rule violates local tomography and maintains a single fermionic mode without ever being in a local superposition of the classical basis. We show how this can be exploited to generate entanglement. In the non-Abelian anyonic example, we improve on the fermionic example by creating entanglement generation while the mediator state remains a pure state, raising interesting questions about the physical meaning of purity in this theory. Thus, no stochasticity is required in the mediator by having a non-Abelian constrained quantum theory. 

However, these two examples both have the shortcoming of only allowing the mediator to have only $2$ or $3$ distinguishable states. Our last example that uses the bit anti-bit toy theory allows the locally classical mediator to have any number of distinguishable classical states. Nevertheless, one shortcoming of this last example is that the encoding of our usual qubits in the toy theory is not as neat as in the two previous examples. 

The results presented in our work are not an attempt to provide a realistic model of gravity-matter interaction. Our goal is to demonstrate that local tomography can be relaxed in the context of the BMV experiment, and still lead to GIE. 

The reader may wonder whether one should consider non-locally tomographic couplings as relevant. We believe one should, especially in the context of quantum gravity. The toy models we proposed are based on standard constrained quantum systems. Especially if one considers carefully the modelling of the gravitational mediator by the classical theory of general relativity and the quantum matter as a theory of quantum field theory, such as QED or the Standard Model. The reason is that all these theories can be considered gauge field theories. A gauged quantum theory takes the form of a constrained quantum theory, giving rise to superselection rules and, thus, to the violation of local tomography \cite{dariano_fermionic_2014}. Furthermore, let us consider that the symmetries and gauges present in general relativity are expected to be extended to the quantum field theory that models the matter sector. It is reasonable to expect and consider structures where the global theory of the coupled dynamical spacetime with the quantum field theory is non-trivially superselected in the quantum sector, and local tomography is not satisfied.  We can even claim that it would be reasonable to expect a situation resembling the non-Abelian anyon example in such a scenario, given the known non-Abelian nature of the gauge constraints present in classical general relativity \cite{blagojevic_gravitation_2002}. Therefore, considering non-local tomography is not just a mathematical or over-idealised possibility. It is well-grounded, as it can reasonably be expected to emerge in the situation described in the BMV thought experiment.    

Let us interpret our results more physically. Let us consider the gravitational mediator as a dynamical spacetime, as in general relativity. We would have a spacetime that, when observed independently, would behave entirely classically, thus possibly following the description of general relativity. With the non-Abelian anyon example coupling, we have shown that spacetime would not even necessarily need stochasticity. Therefore, we can imagine spacetime being locally described by general relativity, having a pure classical state describing it when considering it independently of matter. So, no spacetime superposition observed, no mention of gravitons or self-interfering spacetime.

Let us now consider the properties of the bit anti-bit example. We can even consider that such spacetime remains described by a pure classical state when interacting and probed by "classical" matter. In our example, if the bits of the matter sector interact with the bits of the gravitational sector, such interaction would be entirely classical, and the local state of the gravitational mediator would remain purely classical. So, we have shown that within the BMV thought experiment, it is possible to have locally mediated entanglement generation with a spacetime that appears to be classical when probed by "classical" matter or independently of matter. A classical theory of spacetime, such as general relativity, can describe the local states of spacetime. Nevertheless, such a dynamical spacetime can have a non-locally tomographic coupling to quantum matter, allowing gravitational entanglement generation through system-local mediation.

\section*{Acknowledgements}
This work was supported by the Hong Kong Research Grant Council through the Senior Research Fellowship Scheme SRFS2021-7S02 and the Research Impact Fund R703521F, by the Chinese Ministry of Education through Grant No. 2023ZD0300600, and by the John Templeton Foundation through the ID No. 62312 grant, as part of the “The Quantum Information Structure of Spacetime” project (QISS). The opinions expressed in this publication are those of the authors and do not necessarily reflect the views of the John Templeton Foundation.

\bibliographystyle{apsrev4-1}
\bibliography{reference}

\end{document}